\documentclass[conference,a4paper]{IEEEtran}
\ifCLASSINFOpdf
  \usepackage[pdftex]{graphicx}
\else
  \usepackage[dvips]{graphicx}
\fi
\usepackage[table]{xcolor}
\usepackage{balance}
\usepackage{tabularx}
\usepackage{comment}
\usepackage{amsmath,amssymb}
\ifCLASSOPTIONcompsoc
 \usepackage[caption=false,font=normalsize,labelfont=sf,textfont=sf]{subfig}
\else
 \usepackage[caption=false,font=footnotesize]{subfig}
\fi
\newcommand{\Ss}[0]{\scriptscriptstyle \mathrm{SS}}

\newcommand{\ris}[0]{\scriptscriptstyle \mathrm{RIS}}

\newcommand{\rs}[0]{\scriptscriptstyle \mathrm{RS}}
\newcommand{\ST}[0]{\scriptscriptstyle \mathrm{ST}}
\newcommand{\RT}[0]{\scriptscriptstyle \mathrm{RT}}

\newcommand{\AoA}[0]{\mathrm{AoA}}
\newcommand{\AoD}[0]{\mathrm{AoD}}

\newcommand{\change}[1]{{\color{black}#1}}
\renewcommand{\baselinestretch}{0.977}
\makeatletter
\let\oldabs\abs
\def\abs{\@ifstar{\oldabs}{\oldabs*}}

\let\oldnorm\norm
\def\norm{\@ifstar{\oldnorm}{\oldnorm*}}
\makeatother
\renewcommand{\a}{\mathbf{a}}

\newcommand{\g}{\mathbf{g}}
\newcommand{\h}{\mathbf{h}}

\newcommand{\z}{\mathbf{z}}

\newcommand{\0}{\mathbf{0}}

\newcommand{\I}{\mathbf{I}}

\newcommand{\Z}{\mathbf{Z}}

\newcommand{\Phib}{\mathbf{\Phi}}
\newcommand{\Compl}{\mbox{$\mathbb{C}$}}

\newcommand{\herm}{\mathrm{H}}

\title{Realistic Evaluation of Impedance-Based RIS Modeling: Practical Insights and Applications}

\author{
    \IEEEauthorblockN{Ayane Lebeta Goshu\IEEEauthorrefmark{1},   % 1st author, 1st affiliation
    Placido Mursia\IEEEauthorrefmark{1},   % 2nd author, 2nd affiliation
    Vincenzo Sciancalepore\IEEEauthorrefmark{1},    % author, 2nd affiliation
    Marco Di Renzo\IEEEauthorrefmark{2}\IEEEauthorrefmark{3},      % 7th author, 4th affiliation
    Xavier Costa-Pérez\IEEEauthorrefmark{1}\IEEEauthorrefmark{4}    '/
    }                                     % ...
    \IEEEauthorblockA{\IEEEauthorrefmark{1}% 1st affiliations
    NEC Laboratories Europe GmbH, Heidelberg, Germany}
    \IEEEauthorblockA{\IEEEauthorrefmark{2}% 4th affiliations
    Université Paris-Saclay, CNRS, Centrale Supélec, Gif-sur-Yvette, France} 
    \IEEEauthorblockA{\IEEEauthorrefmark{3}% 5th affiliations
    King's College London, London, United Kingdom}
    \IEEEauthorblockA{\IEEEauthorrefmark{4}% 5th affiliations
    i2CAT Foundation and ICREA, Barcelona, Spain}

}
\begin{document}
\maketitle

\begin{abstract}
Reconfigurable Intelligent Surfaces (RISs) have emerged as a promising technology for next-generation wireless communications, offering energy-efficient control of electromagnetic (EM) waves. While conventional RIS models based on phase shifts and amplitude adjustments have been widely studied, they overlook complex EM phenomena such as mutual coupling, which are crucial for advanced wave manipulations. Recent efforts in EM-consistent modelling have provided more accurate representations of RIS behavior, highlighting challenges like structural scattering—an unwanted signal reflection that can lead to interference. In this paper, we analyze the impact of structural scattering in RIS architectures and compare traditional and EM-consistent models through full-wave simulations, thus providing practical insights on the realistic performance of current RIS designs. Our findings reveal the limitations of current modelling approaches in mitigating this issue, underscoring the need for new optimization strategies.
\end{abstract}

\vskip0.5\baselineskip
\begin{IEEEkeywords}
 RIS, mutual coupling, structural scaterring, antennas, electromagnetic, propagation, full-wave simulations.
\end{IEEEkeywords}

\let\thefootnote\relax\footnotetext{Ayane Lebeta Goshu, Placido Mursia and Vincenzo Sciancalepore are with the 6G Networks group, NEC Laboratories Europe GmbH, Heidelberg, Germany (Email: \{name.surname\}@neclab.eu). Marco Di Renzo is with Université Paris-Saclay, CNRS, CentraleSupélec, Laboratoire des Signaux et Systèmes, 3 Rue Joliot-Curie, 91192 Gif-sur-Yvette, France. (Email: marco.di-renzo@universite-paris-saclay.fr), and with King's College London, Centre for Telecommunications Research -- Department of Engineering, WC2R 2LS London, United Kingdom (Email: marco.di\_renzo@kcl.ac.uk). Xavier Costa-Pérez is with i2CAT Foundation and ICREA, Barcelona, Spain (Email: xavier.costa@ieee.org).
}

%%%%%%%%%%%%%%%%%%%%%%%%%%%%%%%%%%%%%%%%%%%%%%%%%%%%%%%%%%%%%%%%%%%%%%%%%%
\section{Introduction}
%%%%%%%%%%%%%%%%%%%%%%%%%%%%%%%%%%%%%%%%%%%%%%%%%%%%%%%%%%%%%%%%%%%%%%%%%%

Reconfigurable Intelligent Surfaces (RISs) are gaining attention as a key technology for next-generation wireless communications, offering a new paradigm where the wireless channel becomes a programmable component. They are engineered materials designed with nearly passive unit cells that control RF signals, improve coverage, enhance connectivity in non-line-of-sight scenarios, and manipulate electromagnetic (EM) waves~\cite{Basar24}. Indeed, RISs operate without power-hungry electronics, reducing energy consumption and hardware costs, making them an efficient solution for boosting energy efficiency in future wireless networks, as well as enabling novel applications such as integrated sensing and communications (ISAC)~\cite{Strinati24}. In this regard, to date, there exists several RIS prototypes that empirically demonstrate its capabilities and practical feasibility (see, e.g.,~\cite{Rossanese2024, Gros21}).

A key component to unlock the potential of the RIS technology lies in accurate modelling of its EM behavior. RIS optimization under conventional channel modelling, which is based on phase shifts and amplitude attenuation across the RIS, is well understood, and it has been exploited for several applications ranging from energy efficiency~\cite{Yang22} to ISAC~\cite{Xing23}. However, such approaches fail to describe important EM effects, such as the mutual coupling, and therefore do not allow obtaining advanced wave manipulations. Therefore, recently, there has been a growing interest in EM-consistent modelling of RIS~\cite{Alexandropoulos21}. In~\cite{Gradoni21}, an initial EM-consistent and mutual coupling aware RIS channel model based on mutual impedances was proposed for the case of pure line-of-sight (LoS) propagation. This was then extended to account for the presence of scattering objects in~\cite{Mursia23}, and experimentally proven to provide accurate modelling of realistic propagation environments in~\cite{Mursia24}. Whereas, the authors of~\cite{Nerini24} have generalized such model by using both impedances and scattering parameters, which are inherently equivalent, but exhibit differences in terms of mathematical tractability. In~\cite{Pettanice24}, such model has been validated via full-wave simulations. In addition, other EM-consistent RIS modelling approaches include characterizing the EM response of the RIS via the polarizability of its elements~\cite{Faqiri23}. Such intense study on accurate EM modelling of RISs has led to innovative applications, such as the design of conformal surfaces~\cite{Mursia24_J}.

Unlike conventional RIS propagation models, the aforementioned EM-consistent approaches highlight an often overlooked aspect, which is related to unwanted and uncontrollable scattering caused by the structure of the RIS itself, dubbed as \emph{structural scattering}~\cite{Abrardo24}. If not properly accounted for, such effect has been shown to produce an unwanted signal reflection along the specular direction with respect to the incoming wave, with a maximum intensity comparable to that of the intended beam steering, thereby potentially causing substantial interference, security issues, and inaccuracies in localization and sensing procedures.

In this paper, we shed some light on the practical impact of the structural scattering produced by a conventional RIS architecture based on PIN diodes in the millimeter-wave range. Specifically, we compare two alternative strategies to optimize the RIS configuration, namely based on conventional and EM-consistent models, respectively, and provide full-wave simulations of the RIS response. While our findings demonstrate the severity of the RIS structural scattering, which may cause significant interference, we argue that a careful network planning may partially overcome such limitation. Moreover, numerical results demonstrate that both modelling approaches fail to alleviate the structural scattering, thereby motivating the need for novel advanced RIS design and optimization methods.

\textbf{Notation.} We use bold font lower and upper case for vectors and matrices, respectively. %\Im\{\cdot\}$, $\Re\{\cdot\}$ stand for the imaginary and real part, respectively. 
$\0$ represents the all-zero matrix, $(\cdot)^\herm$ denotes the hermitian transpose operator, while $j=\sqrt{-1}$ is the imaginary number. $\otimes$ represents the Kronecker product, while $|x|$ and $\angle x$ denote the absolute value and phase of the complex number $x$, respectively.

%%%%%%%%%%%%%%%%%%%%%%%%%%%%%%%%%%%%%%%%%%%%%%%%%%%%%%%%%%%%%%%%%%%%%%%%%%
\section{RIS channel modelling and optimization}
%%%%%%%%%%%%%%%%%%%%%%%%%%%%%%%%%%%%%%%%%%%%%%%%%%%%%%%%%%%%%%%%%%%%%%%%%%

Our objective is to highlight the impact of structural scattering produced by a commonly-used RIS hardware. Therefore, in order to highlight this issue, we consider a single-input single-output (SISO) scenario, wherein a single-antenna transmitter (TX) communicates with a single-antenna receiver (RX) via an $N$-element planar RIS, which is equipped with $N_x$ and $N_z$ elements along the $x$- and $z$-axis, respectively, with $N=N_x\times N_z$. Furthermore, we assume that all three devices are in the far-field of each other and that the direct link between the TX and the RX is blocked. In particular, the transmit signal impinges on the RIS from an angle-of-arrival (AoA) denoted as $\AoA^{a}$, $\AoA^{e}$ along the azimuth and elevation directions, respectively. Similarly, the reflected signal departs the RIS along an angle-of-departure (AoD) denoted as $\AoD^{a}$, $\AoD^{e}$ along the azimuth and elevation directions, respectively.
In the following, we briefly review two widely-used alternative RIS modelling and associated optimization methods, in order to perform a rigorous comparison in terms of RIS EM response, as described in Section~\ref{sec:results}.

%%%%%%%%%%%%%%%%%%%%%%%%%%%%%%%%%%%%%%%%%%%%%%%%%%%%%%%%%%%%%%%%%%%%%%%%%%
\subsection{Conventional channel model}\label{subsec:conv_ch}
%%%%%%%%%%%%%%%%%%%%%%%%%%%%%%%%%%%%%%%%%%%%%%%%%%%%%%%%%%%%%%%%%%%%%%%%%%
Let $\g\in \Compl^{N\times 1}$ and $\h\in \Compl^{N\times 1}$ denote the channels from the TX to the RIS, and from the RIS to the RX, respectively. The received signal at the RX is given by
\begin{align}
    y = \h^\herm \Phib \g x + n \quad \in \Compl,
\end{align}
where $\Phib = \mathrm{diag}(e^{j\phi_1},\ldots,e^{j\phi_N})\in\Compl^{N\times N}$ denotes the matrix of phase shifts at the RIS, $x\in\Compl$ stands for the transmit signal and $n$ is the noise coefficient, which is distributed as $\mathcal{CN}(0,\sigma_n^2)$. Given the planar structure of the considered RIS, we define the 2D array response as
\begin{align}
    \a({\theta},{\phi}) & \triangleq \a_x({\theta}) \otimes \a_z({\phi})\label{eq:UPA} \\
     & = [1,e^{j2\pi\delta \cos({\theta})},\ldots,e^{j2\pi\delta(N_x-1) \cos({\theta})}]  \nonumber\\
     & \otimes [1,e^{j2\pi\delta \sin({\phi})},\ldots,e^{j2\pi\delta(N_z-1) \sin({\phi})}] \in \Compl^{N\times 1} ,
\end{align}
where $\theta,\,\phi$ represent the azimuth and elevation steering angles, respectively, and $\delta=d/\lambda$ stands for the ratio between the antenna spacing $d$ and the signal wavelength $\lambda$. Therefore, we have
\begin{align}
    \h & = \sqrt{\gamma_h} \, \a({\AoD^a},{\AoD^e})\\
    \g & = \sqrt{\gamma_g} \, \a({\AoA^a},{\AoA^e}),
\end{align}
where $\gamma_h$ and $\gamma_g$ denote the pathloss of the two channel links, respectively.

Under this channel modelling, it is well-known that the RIS configuration maximizing the received signal-to-noise ratio (SNR) at the RX is given by
\begin{align}\label{eq:phi_opt}
    \phi_n^\star = \angle g_n - \angle h_n,\quad n=1,\ldots,N.
\end{align}

%%%%%%%%%%%%%%%%%%%%%%%%%%%%%%%%%%%%%%%%%%%%%%%%%%%%%%%%%%%%%%%%%%%%%%%%%%
\subsection{Impedance-based model}\label{subsec:Z_model}
%%%%%%%%%%%%%%%%%%%%%%%%%%%%%%%%%%%%%%%%%%%%%%%%%%%%%%%%%%%%%%%%%%%%%%%%%%

Following the approach in~\cite{Gradoni21}, we model all the antennas in the scenario as loaded wire dipoles. Therefore, the end-to-end channel from the TX to the RX via the RIS is expressed via the self and mutual impedances as
\begin{align}\label{eq:H}
    \mathcal{H} = y_0 \big[Z_{\RT}-\z_{\rs}\big(\Z_{\Ss}+\Z_{\ris}\big)^{-1}\z_{\ST}\big] \in \Compl,
\end{align}
where $y_0\in\Compl$ is a constant that accounts for the internal losses and loads of the TX and the RX, while the subscripts $\{\mathrm{R,T,S}\}$ denote the RX, the TX, and the RIS, respectively. Here, $Z_{\RT}=0$ denotes the mutual impedance between the TX and the RX, which is assumed to be equal to zero due to the considered NLoS scenario, while $\z_{\rs}\in\Compl^{1\times N}$ and $\z_{\ST}\in\Compl^{N\times 1}$ denote the mutual impedance between the RIS and the RX, and between the TX and the RIS, respectively. $\Z_{\Ss}\in\Compl^{N\times N}$ denotes the self and mutual impedances among the unit cells of the RIS, while $\Z_{\ris}\in\Compl^{N\times N}$ is a diagonal matrix containing the tunable loads at the RIS elements, which is modeled as
\begin{align}
    Z_{\ris}(n,n) = R_0 + j X_n \quad n=1,\ldots,N.\label{eq:Zris}
\end{align}
In Eq.~\eqref{eq:Zris}, $R_0$ and $X_n$ denote the resistance and tunable reactance of the $n$-th RIS element load. Also, we assume that $\Z_{\Ss}$ is a diagonal matrix with elements $Z_{\Ss}(n,n) = X_{\Ss} + j Y_{\Ss}$ $\forall n$. Lastly, we remark that all self and mutual impedances in Eq.~\eqref{eq:H} can be obtained as described in~\cite{Gradoni21, Mursia23}.

Following the approach in~\cite{Qian2020}, let $a_n = \frac{z_{\ST}(n)z_{\rs}(n)}{2|R_0+X_{\Ss}|}$ and $b = Z_{\RT}-\sum_n a_n$. Hence, under the assumption that the mutual coupling among the RIS elements is negligible, the RIS configuration that maximizes the SNR at the RX is given by
\begin{align}\label{eq:Z_opt}
    Z_{\ris}^\star(n,n) = \frac{2|R_0 + X_{\Ss}|}{1+e^{j2\theta^\star_n}} - Z_{\Ss}(n,n) \quad \forall n,
\end{align}
where we have defined
\begin{align}
    2\theta^\star_n = \angle b - \angle a_n \quad \forall n.
\end{align}

%%%%%%%%%%%%%%%%%%%%%%%%%%%%%%%%%%%%%%%%%%%%%%%%%%%%%%%%%%%%%%%%%%%%%%%%%%
\subsection{Structural scattering}
%%%%%%%%%%%%%%%%%%%%%%%%%%%%%%%%%%%%%%%%%%%%%%%%%%%%%%%%%%%%%%%%%%%%%%%%%%

The EM-consistent model in Eq.~\eqref{eq:H}, reveals an interesting insight. Indeed, as pointed out in~\cite{Abrardo24}, even when the reflection coefficient of the RIS is set to zero, i.e., $\Z_{\ris} = Z_0 \I$, such that the tunable loads of the RIS are matched to the reference impedance $Z_0$ and do not re-radiate the incoming signal, 

%even in the absence of tunable loads at the RIS, i.e., $\Z_{\ris} = \0$, 
the channel coefficient in the considered NLoS scenario is non-zero and given by
\begin{align}\label{eq:H_ss}
    \mathcal{H} = -y_0 \z_{\rs}(\Z_{\Ss}+Z_0 \I)^{-1}\z_{\ST},
\end{align}
which is denoted as \emph{structural scattering}. Eq.~\eqref{eq:H_ss} represents an unwanted and uncontrollable signal scattering from the RIS, which is due to its physical structure, and is always present, regardless of whether the channel is LoS or not, and even when the RIS reflection coefficient is zero. 

In the following, we detail a practical RIS array design based on commonly-used components, and provide realistic full-wave simulations of its EM response based on both RIS channel models presented in Sections~\ref{subsec:conv_ch} and~\ref{subsec:Z_model} with the corresponding optimization strategies in Eqs.~\eqref{eq:phi_opt} and~\eqref{eq:Z_opt}, respectively. In particular, we highlight the practical impact of the structural scattering on the overall RIS performance.

%%%%%%%%%%%%%%%%%%%%%%%%%%%%%%%%%%%%%%%%%%%%%%%%%%%%%%%%%%%%%%%%%%%%%%%%%%
\section{RIS design}\label{sec:array}
%%%%%%%%%%%%%%%%%%%%%%%%%%%%%%%%%%%%%%%%%%%%%%%%%%%%%%%%%%%%%%%%%%%%%%%%%%

\begin{figure}[!htbp]
    \centering \includegraphics[width=0.25\textwidth]{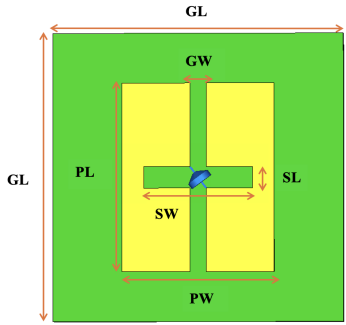} 
    \caption{Considered RIS unit cell structure.}
    \label{fig:Unit_cell}
\end{figure}

%%%%%%%%%%%%%%%%%%%%%%%%%%%%%%%%%%%%%%%%%%%%%%%%%%%%%%%%%%%%%%%%%%%%%%%%%%
%\subsection{Unit cell design}\label{sec:sm}
%%%%%%%%%%%%%%%%%%%%%%%%%%%%%%%%%%%%%%%%%%%%%%%%%%%%%%%%%%%%%%%%%%%%%%%%%%
As illustrated in Fig.~\ref{fig:Unit_cell}, the unit cell is constructed from two copper patches separated by a gap, with a horizontal slot located at the center. To enable a phase change in the reflected wave, a PIN diode (MADP-000907) model from MACOM is placed between the patches~\cite{PIN}. The dielectric substrate used is Rogers RT/duroid 5880, characterized by a permittivity of $2.2$, a loss tangent of $0.0009$ at $10$~GHz, and a thickness of $0.5$~mm. This unit cell is specifically designed to resonate at an operating frequency of $28$~GHz, with a dimension of $\lambda$/2.

%Floquet ports are employed to excite the unit cell structure, enabling the analysis of a single element while minimizing computational complexity of full-wave simulations using CST~\cite{cst}. By switching the diode between OFF and ON states, the resonant frequency of the unit cell is modified, achieving a $1$-bit reflection phase. %This allows precise control over the reflection characteristics of the RIS, enabling effective beam steering.
Floquet ports are employed to excite the unit cell structure, allowing for the analysis of a single element while reducing the computational complexity of full-wave simulations in CST~\cite{cst}. %By toggling the diode between its OFF and ON states, the unit cell's resonant frequency is altered, enabling a $1$-bit reflection phase shift. 
Since CST does not have a direct model for a PIN diode, we simulate its behavior using lumped elements in a series resistor (R), inductor (L), and capacitor (C) configuration. In the ON state, the PIN diode is represented by R and L in series, mimicking its conductive property. In the OFF state, the model consists of L and C, capturing the diode’s capacitive characteristics when non-conductive.

\begin{figure}[!t]
    \centering \includegraphics[width=0.45\textwidth]{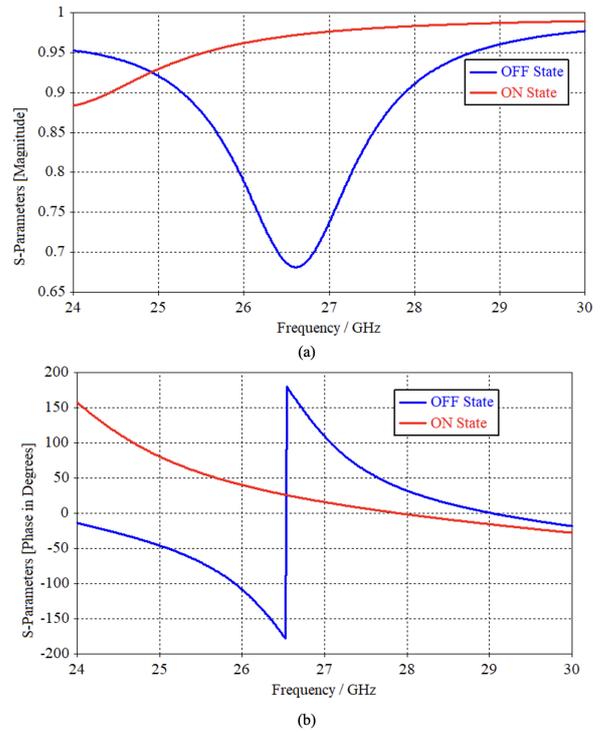} 
    \caption{Unit cell reflection coefficient (a) Reflection magnitude (b) Reflection phase.}
    \label{fig:Reflection_coeffiecient}
\end{figure}

Fig.~\ref{fig:Reflection_coeffiecient} illustrates the reflection coefficient performance of the proposed unit cell design with the PIN switches in both ON and OFF states. The unit cell achieves a balanced performance in phase difference and reflection magnitude at a frequency of $26.168$~GHz.
%At the operating frequency of 28 GHz, the unit cell experiences good reflection performance but insufficient phase difference between the two states. Therefore, we select the operating frequency of $26.168$~GHz, in order to achieve a balanced performance in terms of both phase difference and reflection magnitude.

%%%%%%%%%%%%%%%%%%%%%%%%%%%%%%%%%%%%%%%%%%%%%%%%%%%%%%%%%%%%%%%%%%%%%%%%%%
%\subsection{Array Design}
%%%%%%%%%%%%%%%%%%%%%%%%%%%%%%%%%%%%%%%%%%%%%%%%%%%%%%%%%%%%%%%%%%%%%%%%%%

Leveraging the previously described unit cell design, we develop an RIS consisting of $20 \times 20$ unit cell elements, whose EM response is detailed next.
%. To assess the influence of different configurations on the reflected beam direction and overall performance, we conducted a detailed analysis of the RIS array. In this study, two scenarios were considered for exciting the RIS surface: normal incidence and oblique incidence of a plane wave.

%%%%%%%%%%%%%%%%%%%%%%%%%%%%%%%%%%%%%%%%%%%%%%%%%%%%%%%%%%%%%%%%%%%%%%%%%%
\section{Numerical results and discussion}\label{sec:results}
%%%%%%%%%%%%%%%%%%%%%%%%%%%%%%%%%%%%%%%%%%%%%%%%%%%%%%%%%%%%%%%%%%%%%%%%%%

In order to assess the performance of the RIS design detailed in Section~\ref{sec:array}, we evaluate its radar cross-section (RCS) under plane wave illumination using full-wave simulations in CST. The state of each RIS unit cell is optimized according to the analytical frameworks described in Sections~\ref{subsec:conv_ch} and~\ref{subsec:Z_model}. As for the channel model in Section II-A, since the PIN diode-based RIS has a $1$-bit phase-shifting capability, we quantize each optimized phase shift in Eq.~\eqref{eq:phi_opt} with the closest feasible point. As for the channel model in Section II-B, we model the tunable load of each unit cell as an equivalent RLC circuit as
\begin{align}
    Z_{\ris}(n,n) = R_0 + j 2 \pi f L + \frac{1}{j2\pi f C} \quad \forall n,
\end{align}
where $f$ is the working frequency, $L$ is the load inductance and $C$ is its tunable capacitance. As described in Section~\ref{sec:array}, the phase-shifting is provided by a commercially-available lumped element whose capacitance ranges between $C_{\mathrm{min}}=0.025$~pF and $C_{\mathrm{max}}=0.03$~pF. Therefore, we quantize each optimized RIS reactance in Eq.~\eqref{eq:Z_opt} to fall within this range. All relevant simulation parameters are reported in Table~\ref{tab:sim_params}.

\begin{table}[t!]
\caption{Simulation parameters.}
\label{tab:sim_params}
\centering
\resizebox{1\linewidth}{!}{%
\renewcommand{\arraystretch}{1.0}
\begin{tabular}{|c|c|c|c|c|c|}
\hline
\cellcolor[HTML]{EFEFEF} \textbf{Parameter} & \textbf{Value} & \cellcolor[HTML]{EFEFEF} \textbf{Parameter} & \textbf{Value} & \cellcolor[HTML]{EFEFEF} \textbf{Parameter} & \textbf{Value}\\
\hline
 \cellcolor[HTML]{EFEFEF} $f$ & $26.168$~GHz & \cellcolor[HTML]{EFEFEF} $d/\lambda$ & $0.5$ & \cellcolor[HTML]{EFEFEF} $N_x$ & $20$ \\
\hline 
 \cellcolor[HTML]{EFEFEF} $N_y$ & $20$ & \cellcolor[HTML]{EFEFEF} $\AoA^a$ & $[\change{-30,\,30]^\circ}$ & \cellcolor[HTML]{EFEFEF} $\AoD^a$ & $[-30,\,\change{45]^\circ}$ \\
\hline 
 \cellcolor[HTML]{EFEFEF} $\AoA^e$ & $[0,\,-60]^\circ$ & \cellcolor[HTML]{EFEFEF} $\AoD^e$ & $0^\circ$ & \cellcolor[HTML]{EFEFEF} $R_0$ & $5.2$~$\Omega$  \\
 \hline
 \cellcolor[HTML]{EFEFEF} $GL$ & $5.35$~mm & \cellcolor[HTML]{EFEFEF} $SW$ & $2$~mm & 
 \cellcolor[HTML]{EFEFEF} $PL$ & $3.5$~mm \\
 \hline
 \cellcolor[HTML]{EFEFEF} $SL$ & $0.4$~mm & 
 \cellcolor[HTML]{EFEFEF} $PW$ & $2.8$~mm &
  \cellcolor[HTML]{EFEFEF} $GW$ & $0.3$~mm \\
\hline
 \cellcolor[HTML]{EFEFEF} $L$ & $30$~pH & 
 \cellcolor[HTML]{EFEFEF} $C_{\mathrm{min}}$ & $0.025$~pF &
  \cellcolor[HTML]{EFEFEF} $C_{\mathrm{max}}$ & $0.03$~pF \\
\hline
\end{tabular}%
}
\renewcommand{\arraystretch}{1}
\end{table}

%%%%%%%%%%%%%%%%%%%%%%%%%%%%%%%%%%%%%%%%%%%%%%%%%%%%%%%%%%%%%%%%%%%%%%%%%%
\subsection{Conventional channel model}\label{subsec:results_conv}
%%%%%%%%%%%%%%%%%%%%%%%%%%%%%%%%%%%%%%%%%%%%%%%%%%%%%%%%%%%%%%%%%%%%%%%%%%
\begin{figure}[!t]
    \centering
    \includegraphics[width=0.48\textwidth]{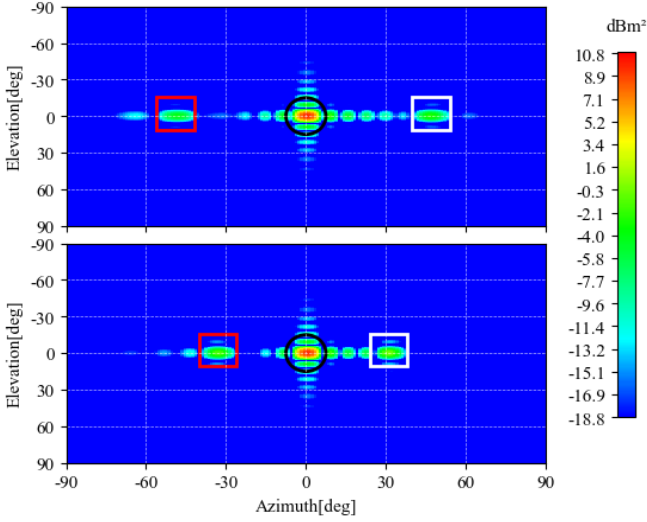} 
    \caption{RIS RCS obtained via full-wave simulation for the case of conventional channel modelling, $\AoA^e = \AoD^e = 0^\circ$ along the elevation, and $\AoA^a = 0^\circ$, $\AoD^a = \change{45^\circ}$ (upper-part) and $\AoA^a = 0^\circ$, $\AoD^a = \change{30^\circ}$ (lower-part) along the azimuth.}
    \label{fig:zero_30_45}
\end{figure}

\begin{figure}[!t]
    \centering
    \includegraphics[width=0.48\textwidth]{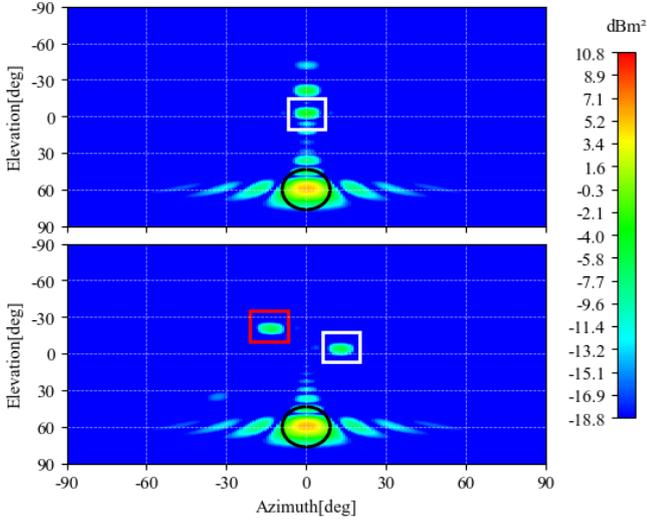} 
    \caption{RIS RCS obtained via full-wave simulation for the case of conventional channel modelling, $\AoA^e = -60^\circ$, $\AoD^e = 0^\circ$ along the elevation, $\AoA^a = -30^\circ$, $\AoD^a = 30^\circ$ (upper-part) and $\AoA^a = -30^\circ$, $\AoD^a = \change{45^\circ}$ (lower-part) along the azimuth.}
    \label{fig:thirty_30_45}
\end{figure} 

In Fig.~\ref{fig:zero_30_45}, we show the RIS RCS for the case of normal incidence along both azimuth and elevation, i.e., $\AoA^e = \AoA^a = 0^\circ$, while the RIS configuration is optimized to achieve steering towards $\AoD^e = 0^\circ$, $\AoD^a = \change{45^\circ}$ (upper part), and $\AoD^e = 0^\circ$, $\AoD^a = \change{30^\circ}$ (lower part). We notice that in both cases, the RIS achieves accurate beam steering along the azimuth direction (indicated with \change{white box}), although the low-resolution of the PIN diodes results in a symmetric beam as compared to the normal to the surface \change{(red box)}. \change{The low-resolution of the considered PIN diodes introduces a quantization effect that results in an undesired beam towards the direction that is symmetrical to the intended one, adversely affecting the performance of practical RIS implementations~\cite{DIRENZO2023}}. However, the structural scattering lobe (indicated with a \change{black} circle), which is located along the specular direction with respect to the incoming wave, i.e., at the center of the plot, exhibits a magnitude of approximately $10.7$~\change{dBm$^2$}, as compared to $-1$~\change{dBm$^2$} of the \change{intended} lobes. Therefore, we conclude that when the RX is positioned at the same elevation of the RIS, the structural scattering provides substantial unwanted interference, which may limit the system performance, produce inaccurate or false sensing results, and cause security issues in the case of eavesdroppers.

Fig.~\ref{fig:thirty_30_45} depicts how this issue is partially mitigated by proper network deployment. Indeed, in this case, the incoming wave arrives with an angle of $\AoA^e = -60^\circ$, $\AoA^a = -30^\circ$ while the RIS configuration is set to produce the intended reflection in the same direction as the previous case, i.e., $\AoD^e = 0^\circ$ and $\AoD^a = 30^\circ$ (upper-part), and $\AoD^e = 0^\circ$, $\AoD^a = \change{45^\circ}$ (lower-part). Here, the intended lobe (\change{white} box) achieves a RCS of approximately $-4$~\change{dBm$^2$}, while the structural scattering lobe (\change{black} circle) appears to be significantly degraded at $6.2$~\change{dBm$^2$}. Therefore, by suitably tilting the RIS device and/or correctly defining the elevation difference between the TX and the RIS, it is possible to minimize any unwanted interference towards nearby RXs and/or sensing targets.

\begin{comment}
    \begin{figure}[!htbp]
    \centering
    \includegraphics[width=0.48\textwidth]{Figures/0 30 in x axis.pdf} 
    \caption{RIS RCS obtained via full-wave simulation for the case of $\AoA = 0^\circ$, $\AoD = 30^\circ$.}
    \label{fig:zero_30}
\end{figure} 

\begin{figure}[!htbp]
    \centering
    \includegraphics[width=0.48\textwidth]{Figures/0 to 45 in x axis.pdf} 
    \caption{RIS RCS obtained via full-wave simulation for the case of $\AoA = 0^\circ$, $\AoD = 45^\circ$.}
    \label{fig:zero_45}
\end{figure} 
\end{comment}

\begin{comment}
    \begin{figure}[!htbp]
    \centering
    \includegraphics[width=0.48\textwidth]{Figures/-30 -45.pdf} 
    \caption{RIS RCS obtained via full-wave simulation for the case of $\AoA = 30^\circ$, $\AoD = 45^\circ$ and inclined RIS.}
    \label{fig:thirty_45}
\end{figure} 

\begin{figure}[!htbp]
    \centering
    \includegraphics[width=0.48\textwidth]{Figures/-30 -30.pdf} 
    \caption{RIS RCS obtained via full-wave simulation for the case of $\AoA = 30^\circ$, $\AoD = 30^\circ$ and inclined RIS.}
    \label{fig:thirty_30}
\end{figure} 
\end{comment}

\begin{figure}[!t]
    \centering
    \includegraphics[width=0.48\textwidth]{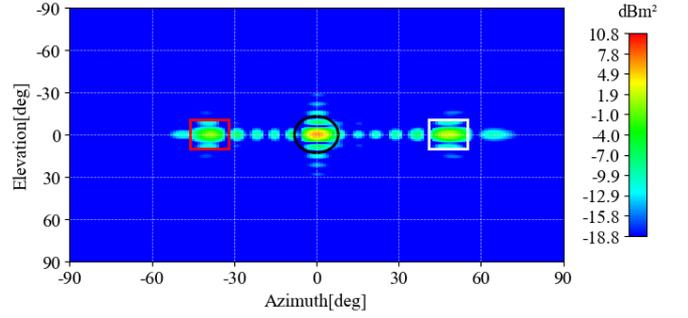} 
    \caption{RIS RCS obtained via full-wave simulation for the case of impedance-based channel modelling, $\AoA^e = \AoD^e = 0^\circ$ along the elevation, and $\AoA^a = 0^\circ$, $\AoD^a = 45^\circ$ along the azimuth.}
    \label{fig:zero_45_Zris}
\end{figure} 

\begin{figure}[!t]
    \centering
    \includegraphics[width=0.48\textwidth]{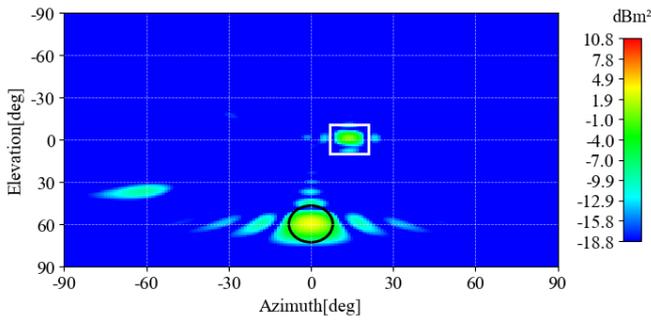} 
    \caption{RIS RCS obtained via full-wave simulation for the case of impedance-based channel modelling, $\AoA^e = -60^\circ$, $\AoD^e = 0^\circ$ along the elevation, $\AoA^a = -30^\circ$, \change{$\AoD^a = 45^\circ$} along the azimuth.}
    \label{fig:thirty_45_Zris}
\end{figure}

%%%%%%%%%%%%%%%%%%%%%%%%%%%%%%%%%%%%%%%%%%%%%%%%%%%%%%%%%%%%%%%%%%%%%%%%%%
\subsection{Impedance-based model}
%%%%%%%%%%%%%%%%%%%%%%%%%%%%%%%%%%%%%%%%%%%%%%%%%%%%%%%%%%%%%%%%%%%%%%%%%%

In this section, we repeat a subset of the experiments detailed in Section~\ref{subsec:results_conv} by exploiting the EM-consistent RIS modelling and optimization strategy described in Section~\ref{subsec:Z_model}.

In Fig.~\ref{fig:zero_45_Zris}, we observe that, compared to the conventional RIS optimization method, the intended lobe \change{(white box)} achieves a higher RCS of $2$~\change{dBm$^2$} and, most importantly, the structural scattering lobe (\change{black} circle) exhibits a smaller dispersion in the angular space, despite having roughly the same maximum RCS. The advantage of EM-compliant RIS optimization is even more evident in Fig.~\ref{fig:thirty_45_Zris}, where the intended lobes exhibit $-2$~\change{dBm$^2$} RCS as compared to $6$~\change{dBm$^2$} of the structural scattering, which is directed away from the RX at an elevation of $60^\circ$. 

However, we remark that in both considered RIS modelling approaches, the structural scattering lobe represents a critical unwanted effect that is not directly taken into account in the optimization of the RIS configuration. Therefore, our findings motivate the need for both careful network planning, and for novel strategies jointly targeting the maximization of the received signal power while limiting the structural scattering effect, as recently elaborated in~\cite{Abrardo24}.

%%%%%%%%%%%%%%%%%%%%%%%%%%%%%%%%%%%%%%%%%%%%%%%%%%%%%%%%%%%%%%%%%%%%%%%%%%
\section{Conclusion}
%%%%%%%%%%%%%%%%%%%%%%%%%%%%%%%%%%%%%%%%%%%%%%%%%%%%%%%%%%%%%%%%%%%%%%%%%%

In this paper, we have provided practical insights on the performance of existing RIS array designs via realistic full-wave simulations. In particular, we have compared both conventional and impedance-based modelling and corresponding optimization methods for RIS configuration by evaluating their corresponding RCS in different physical scenarios. We have argued that current RIS modelling and optimization methods often overlook the \emph{structural scattering}, which is an unintended signal reflection along the specular direction as compared to the incoming wave that may ultimately limit RIS performance if not properly accounted for, resulting in interference, inaccurate sensing results, and security issues.

%%%%%%%%%%%%%%%%%%%%%%%%%%%%%%%%%%%%%%%%%%%%%%%%%%%%%%%%%%%%%%%%%%%%%%%%%%
\section*{Acknowledgment}
%%%%%%%%%%%%%%%%%%%%%%%%%%%%%%%%%%%%%%%%%%%%%%%%%%%%%%%%%%%%%%%%%%%%%%%%%%

This work was supported in part by the MSCA INTEGRATE, and the SNS JU 6G-DISAC Horizon Europe projects under Grant Agreement numbers 101072924 and 101139130, respectively.

\bibliographystyle{IEEEtran}
\bibliography{refs}

\end{document}